\documentclass[prl,amsmath,aps,twocolumn,superscriptaddress]{revtex4}
\usepackage{graphicx,xspace}
\usepackage{float}
\usepackage{amsmath}
\usepackage{amssymb}
\usepackage[normalem]{ulem}
\input{epsf}
\begin{document}
\newcommand{\scalar}[2]{\left \langle#1\ #2\right \rangle}
\newcommand{\me}{\mathrm{e}}
\newcommand{\mi}{\mathrm{i}}
\newcommand{\dif}{\mathrm{d}}
\newcommand{\period}{\text{per}}
\newcommand{\free}{\text{fr}}
\newcommand{\mq}[2]{\uwave{#1}\marginpar{#2}} 
\include{latexcommands}

\title{Moore-Read Fractional Quantum Hall wavefunctions and $SU(2)$ quiver gauge theories}
\author{Raoul Santachiara} \email{raoul.santachiara@u-psud.fr}
\affiliation{CNRS, LPTMS 
             Universit\'e Paris-Sud,UMR 8626, B\^atiment 100, 91405 Orsay, France.} 
\author{Alessandro Tanzini}
\email{tanzini@sissa.it}
\affiliation{SISSA, via Beirut 2-4, 34151, Trieste, Italy.}
  
  \begin{abstract}
We identify Moore-Read wavefunctions, describing non-abelian statistics in fractional quantum Hall systems, with the instanton partition of ${\cal N}=2$ superconformal  quiver gauge theories at suitable values of masses and $\Omega$-background parameters. This is obtained by extending to rational conformal field
theories the $SU(2)$ gauge quiver/Liouville field theory duality recently found by Alday-Gaiotto-Tachikawa.
A direct link between the Moore-Read Hall $n$-body wavefunctions and ${\mathbb Z}_n$-equivariant Donaldson polynomials is pointed out.
  \end{abstract}

\maketitle
{\it Introduction-}
There are now several proposed  experiments aimed at identifying the existence of non-Abelian statistics in nature \cite{Parsa}. The evidence  of such statistics may 
be 
in fact  observed for the first time in fractional quantum Hall (FQH) systems occurring at filling fraction $\nu=5/2$ \cite{Wittell} . 
Non-Abelian phases are gapped phases of matter  in which the adiabatic transport of one excitation around another implies a unitary transformation within a subspace of degenerate wavefunctions which differ from each other only globally \cite{RM_review}.  Using this property it has been shown that systems exhibiting non-Abelian statistics can store topogically protected qubits and are therefore interesting for topological quantum computation \cite{Sarma}.

 Much of the comprehension of  non-Abelian quantum Hall states relies on  the  conformal field theory (CFT) approach \cite{MooreRead,ReadRezayi}. The FQH-CFT connection was suggested by the relation between $(2+1)$ Chern-Simons theory  and rational CFTs \cite{Witten_cscft}.  The corresponding conformal blocks form higher-dimensional representations of the braiding group which can describe quasi-particle statistics in two dimensions. In the CFT approach  \cite{MooreRead,ReadRezayi}, this link  is extended to interpret conformal blocks as  the analytic part of trial wavefunctions for the underlying particles.   The analytic properties of the conformal blocks are  then directly related to the  universal properties characterizing a topological phase, such as the quantum numbers of the ground state and of the excitations. In particular  the effects on  adiabatic exchange  of excitations, and especially  their non-Abelian nature, should be encoded in the monodromy properties of these conformal blocks \cite{RM_review,Read_holonomy}.

In this Letter we consider the Moore-Read (MR) states \cite{MooreRead}. There is   now ample (numerical) evidence that  the physics of the $\nu=5/2$  plateau is well captured  by one particular realization of the MR states which   describes  a (p-wave) pairing of electrons occuring  at  the first excited  Landau level.  The non-Abelian nature of these states is  well understood in terms of  the Ising CFT.
We show that a recent relation found by Alday-Gaiotto-Tachikawa (AGT) between conformal blocks in Liouville field
theory and  instanton partition functions in $SU(2)$ quiver gauge theories
\cite{AGT} can be generalized to the case of conformal blocks of the Ising
CFT.  
A nice geometrical description of ${\cal N}=2$ superconformal quiver gauge theories was provided in \cite{su2quiver} by using M-theory
compactification. It was shown that the data of the quiver are encoded in the extended Teichmuller space of a suitable Riemann
surface with punctures. AGT duality identify the instanton partition function of $SU(2)$ quiver gauge theories with the conformal 
blocks of a Liouville field theory on this Riemann surface. By specializing this duality to Ising CFT, we are able to
provide a direct link with MR wavefunctions.  
As explained in detail later, there is a precise
dictionary between the quantities characterizing  a MR wavefunction and the
ones defining an  $SU(2)$ quiver gauge theory. For a  given number $N_p$ of particles and $n_{qh}$ quasi-holes, we show that:
\textit{i)} there is a corresponding  $SU(2)$ quiver gauge theory with  gauge group $G=\prod_{i}^{N_p+n_{qh}-3} SU(2)_i$ coupled to a total number of $N_t=N_p+n_{qh}$ 
(bi-)fundamental hypermultiplets with a suitable chosen
set of mass parameters; \textit{ii)} the values
of the gauge couplings associated to this quiver gauge theory are fixed by the
positions of the particles and of the quasi-holes; \textit{iii)} each degenerate
quasi-hole wavefunctions is related to a specific Coulomb branch of the quiver gauge theory. 
In this Letter we will give an illustrative and explicit example of this relation by considering the most simple MR wavefunctions exhibiting non-Abelian statistics. 
  
In  earlier works other connections between the FQH and certains (Chern-Simons) effective field theories were mainly  based  on universality arguments \cite{cs_pfaffian}.   
The relation discussed here is instead directly manifest in the specific form of the MR many-body wavefunctions. 
A consequence of our result is that ${\mathbb Z}_n$-equivariant Donaldson polynomials
are encoded in the MR wavefunctions, suggesting the appearance of new unexpected topological features in the FQH.

{\it Moore-Read states-}
In symmetric gauge, the many-body wavefunction $\tilde{\Psi}$ describing $N_p$ particle states in the lowest Landau level  takes the general form:
\begin{equation}
\tilde{\Psi}(\{z_i,\bar{z}_i\})=P_{N_p}(z_1,\cdots,z_N)\prod_{i=1}^{N_p}\mu(z_i,\bar{z}_i),
\end{equation}
where $z_i$ is a complex variable which represents the particle coordinates, $P(z_1,\cdots,z_N)$ is a polynomial in the $N_p$ variables $z_i$  and $\mu(z_i,\bar{z}_i)$, which  is the non-analytic part of the wavefunction, is the measure corresponding to the surface where the particles live.  As $\prod \mu(z_i,\bar{z_i})$ is a simple one body-term , we can drop it for simplifying notations.

\begin{figure}[h]
      \includegraphics[scale=0.45,angle=0]{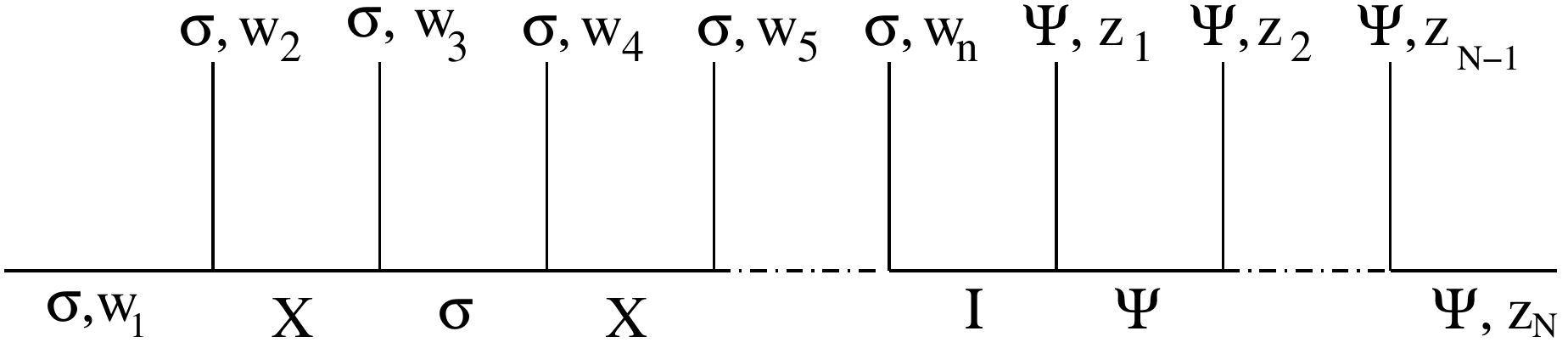}
    \caption{ A diagram representing the  conformal block (\ref{qh_cb}). For the conformal block to be non zero, $n+N$ has to be even. For each diagram there are $n/2-1$ fields $X$ which can correspond to the $Id$ or to the $\sigma$ field, $X=\mbox{Id}$ or $X=\sigma$. The total number of possible conformal block is then $2^{n/2-1}$}  
   \label{cblock}  
   \end{figure}

MR wavefunctions are given by the conformal blocks of the minimal CFT
with central charge $c=1/2$ \cite{diFrancesco}. This is a rational CFT with
two  primary fields, $\Psi$ and $\sigma$, which, together with the identity
$\mbox{Id}$ field, close under the operator algebra. The field $\Psi$, with
conformal dimension $\Delta_{\Psi}=1/2$, is a free fermion field and fuses
with itself into the identity,  $\Psi \times \Psi \to \mbox{Id}$. This field
is identified with the particle operator: the  MR ground state $P_{N_p}(\{z_i\})$ is given by the following   $N_p$- point  correlation function:
\begin{equation}
P_{N_p}^{\mbox{gs}}(\{z_i\})\hat{=}\hspace{-0.05cm}\langle \Psi(z_1) \hdots  \Psi(z_{N_p}) \rangle \prod_{i<j}^{N_p}  z_{ij}^{M+1}=\mbox{Pf}(z^{-1}_{ij}) \prod^{N_p}_{i<j}  z_{ij}^{M+1} , 
\label{gs_mr}
\end{equation}
where $z_{ij}=z_i-z_j$ and $\mbox{Pf}(A_{ij})$ is the Pfaffian of the $N_p\times N_p$ matrix $A_{ij}=z_{ij}^{-1}$. The r.h.s of (\ref{gs_mr}) is obtained by using Wick 
theorem to compute the $N_p-$ free fermions correlation function. For  even (odd) values of $M$,  the state (\ref{gs_mr}) describes bosons (fermions)  
at filling $\nu=1/(1+M)$. 
Hereafter, without any loss of generality,  we can focus on the  bosonic $M=0$ MR states.  
 The field $\sigma$ has conformal dimension  $\Delta_{\sigma}=1/16$ and  represents  the elementary quasi-hole operator.   The subspace of degenerate wavefunctions with the $n_{qh}$ quasi-holes at fixed positions $\{w_i\}$ , $P^{n_{qh}}_{N_p}(\{w_i \},\{z_i\})$, is set by the correlator:
 \begin{eqnarray}
\hspace{-1cm}P^{n_{qh}}_{N_p}(\{w_i \},\{z_i\})&=&\mathcal{F}_{(a)}(\{w_i\},\{z_i\})\times \nonumber \\
&&\times \prod_{i<j}^{N_p} z_{ij} \prod_{i=1}^{N_p}\prod_{j=1}^{n_{qh}} (z_i-w_j)^{\frac{1}{2}} \prod_{i<j}^{n_{qh}} w_{ij}^{\frac{1}{8}}
\label{qh_gen}
\end{eqnarray}
where
\begin{equation}
\mathcal{F}_{(a)}(\{w_i\},\{z_i\})=\langle \sigma(w_1)\cdots\sigma(w_{n_{qh}})\Psi(z_1)\cdots\Psi(z_{N_p})\rangle_{(a)}. 
\label{qh_cb}
\end{equation}
In the above equations  $w_{ij}=w_i-w_j$ while the index $a$ runs over the  the possible conformal blocks. From the fusion between the particle $\Psi$ and the quasi-hole $\sigma$ operators, $\Psi \times \sigma \to \sigma$,  one can verify that 
the particles  and quasi-holes are  mutually local, as required. The dimension of the degenerate space of  the excited wavefunctions (\ref{qh_gen}) is given by the number $2^{n/2-1}$ of conformal blocks corresponding to (\ref{qh_cb}). This number  can be derived from the number of all  possible conformal blocks consistent with the fusions $\sigma \times \sigma= \mbox{Id}+\Psi$ and $\Psi \times \sigma= \sigma$,  as shown in Fig (\ref{cblock}) .

{\it Nekrasov functions and AGT relation}

It has been known for a long time that supersymmetric gauge theories in four dimensions display a topological sector
describing Donaldson polynomials, defined by integration of suitable chiral observables on the instanton moduli space \cite{Witt88}. 
Nekrasov proposed \cite{NekraSW} a deformation of the four-dimensional space-time manifold, the so-called $\Omega$-background, which paved
the way to an explicit evaluation of these integrals via localisation techniques \cite{NekraSW,rubik,BFMT}. The idea is to exploit the rotational
symmetries $U(1)\times U(1)\subset SO(4)$ of $ {\mathbb R}^4 \sim {\mathbb C}^2 $, $(z_1,z_2)\to (e^{i\epsilon_1} z_1, e^{i\epsilon_2} z_2)$
in order to reduce the integral to a discrete sum over
fixed points.
From the mathematical viewpoint, this amounts to consider equivariant Donaldson polynomials \cite{naka}. 
More precisely, Nekrasov's partition function for a $U(r)$ gauge theory
is the generating functional of the integral of the 
fundamental equivariant cohomology class $1\in H^*_T (M_{k,r})$,
where $M_{k,r}$ is the moduli space of $U(r)$ framed instantons with second Chern class $k$, and $T$ the $(r+2)$-dimensional
torus associated to the above mentioned rotational symmetry and to the change of framing, {\it i.e.} $U(1)^r$ global gauge transformations.
In further generality, given two multiplicative classes $E,F$ one defines
\begin{equation}
Z_{inst}^{\epsilon_1,\epsilon_2} (\vec{a},m,\mu; q)= \sum_{k=0}^{\infty} q^k \int_{M_{k,r}} E_{T_m}(T_M) F_{T_\mu}(V)  
\label{zio}
\end{equation}
where $T_M$ is the tangent bundle and $V$ is a vector bundle on $M_{k,r}$, while 
$q^k=e^{2\pi i \tau k}$ is the $k$-instanton action, $\tau$ being the complex coupling of the gauge theory, and
$\epsilon_1,\epsilon_2,\vec{a}= (a_1, \ldots , a_r)$ are the parameters associated to the torus action $T$.
Indeed, in presence of matter multiplets, one has further global symmetries, and one has to include their contribution to the localisation formula. 
For one adjoint hypermultiplet with mass $m$, which has $U(1)_m$ global symmetry, the contribution is given by
the $T_m$ equivariant Euler class of the tangent bundle $E_{T_m}(T_M)$.
For $N_f$ fundamental hypers with masses $\{\mu_f\}= (\mu_1,\ldots,\mu_{N_f})$ one has a $U(N_f)$ global flavour symmetry, and one has
to multiply by the
$T_{{\bf \mu}}$-equivariant Euler class $F_{T_{\bf\mu}}(V)$ associated to the fundamental representation of $U(N_f)$, $T_{{\bf \mu}}$ being its maximal torus
 \cite{BFMT}.

We now propose a mathematical interpretation of the Nekrasov partition functions for quiver gauge theories which are relevant for our problem. 
Quiver gauge theories with gauge group $\prod_{j=1}^n U(r_j)$, $\sum_{j=1}^n r_j = r$ arise naturally
by considering a finite group action ${\mathbb Z}_n$ on the equivariant parameters $a_\alpha \to a_\alpha + 2\pi i \frac{j_\alpha}{n}$,
 $\alpha=1,\ldots,r$.
The integers $r_j$ are the number of times that the $j$-th irreducible representation of the ${\mathbb Z}_n$ group appears in the decomposition.
In particular, in order to obtain the product gauge group of our interest one has to consider $r=2n$ and set $r_j=2$ for any $j$. 
The specialization to $SU(2)_j$ factors come simply by picking the relevant Cartan subgroup $\vec{a}= (a_1, -a_1, \ldots , a_n , -a_n)$.
The quiver has matter fields with masses ${\bf m}= (m_1, \ldots,m_n)$ transforming in the bifundamental $(r_j, \bar r_{j+1})$ representation. The finite group acts 
accordingly on the associated equivariant parameters as $m_j\to m_j + 2\pi i/n$.
In the localization formulae, one needs to restrict only to the fixed point set which is invariant under the finite group action,
namely one consider the ${\mathbb Z}_n$-equivariant classes
 $1^{{\mathbb Z}_n}, E^{{\mathbb Z}_n}_{T_{\bf m}}(T_M)$ associated to the ${\mathbb Z}_n$ action on $M_{k,r}$.
One thus get ${\mathbb Z}_n$-equivariant Donaldson polynomials
\begin{equation}
Z_{inst}^{\epsilon_1,\epsilon_2}(\vec{{\bf a}},{\bf m},{\bf \mu};{\bf q})
= \sum_{{\bf k}=0}^{\infty} {\bf q}^{\bf k} \int_{M_{k,r}} 1^{{\mathbb Z}_n} E^{{\mathbb Z}_n}_{T_{\bf m}}(T_M) 
\prod_{f=1}^4 F_{T_{\mu_f}}(V) 
\label{zion}
\end{equation}
where ${\bf q}^{\bf k}= \prod_{i=1}^n q_i^{k_i}$, $q_i$ encodes the gauge coupling of the $i$-th node and $k_i$ is the associated
second Chern class, with $\sum_i k_i = k$.
\begin{figure}[h]
      \includegraphics[scale=0.45,angle=0]{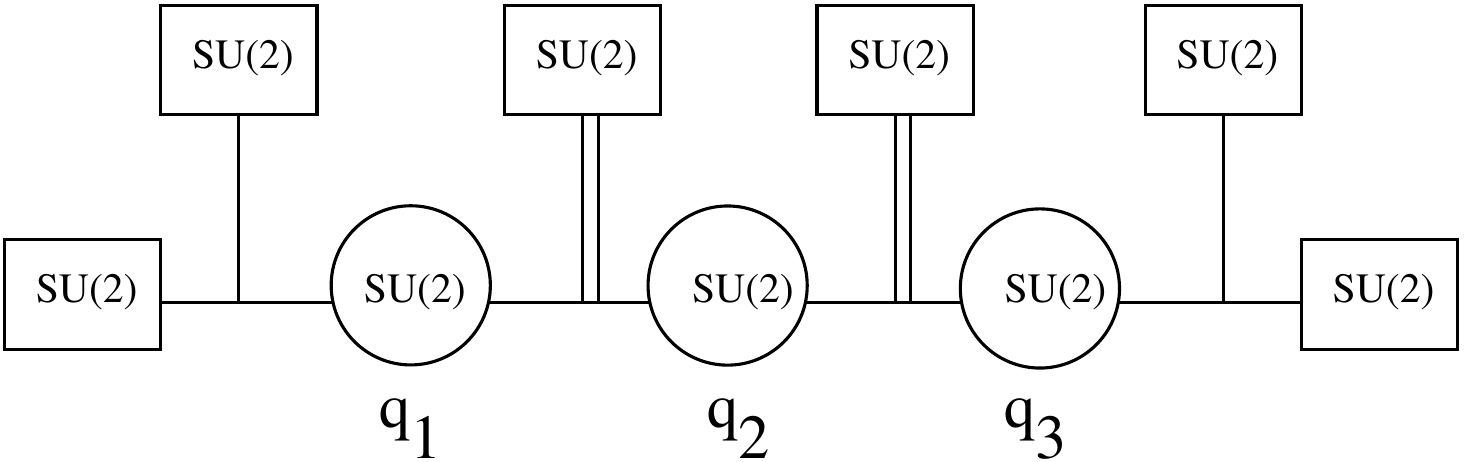}
    \caption{Diagram representing a ${\cal N}=2$ superconformal quivers. The circles corresponds to quiver nodes with $SU(2)$ gauge group.
    The boxes with two outgoing lines represent bifundamental matter and their $SU(2)$ global flavour symmetry. 
    The linear quiver has two fundamental and two antifundamental hypermultiplets at the ends.} 
   \label{quiver}  
   \end{figure}
The contribution $\prod_f F_f$ of the two fundamental and two antifundamental multiplets
at the ends of the linear $SU(2)$ quiver, see Fig.\ref{quiver}, is obtained by imposing appropriate boundary conditions
which constrain the ${\bf k}$-partitions at the final nodes. 
Finally, notice that (\ref{zio}), (\ref{zion}) have an analytic
continuation in the equivariant parameters $\epsilon_1,\epsilon_2,\vec{a}, \mu, {\bf m}$;
this will be relevant in our comparison with the conformal blocks of the Ising model.

In order to spell the precise dictionary of AGT duality, let us recall that
a Liouville field theory is described by a  conformal invariant Lagrangian density
$\mathcal{L}=\frac{1}{4\pi} (\partial \phi)^2 +\lambda e^{2 b \phi}$ where $\lambda$ is some coupling and $b$ can be a general complex number. The   central charge $c$ is given by 
\begin{equation}
c=1+6 Q^2 \quad \mbox{with}\quad Q=b+\frac{1}{b}.
\end{equation}
The basic objects are the exponential fields $e^{\alpha \phi}$ with charge $\alpha$ and conformal dimension 
$\Delta_{\alpha}=\alpha (b+b^{-1}-\alpha)$. Consider a $N$ point correlation function $\langle \prod_{i=1}^{N}e^{\alpha_i \phi(z_i)}\rangle$.  Under  a conformal map, one can send $z_1\to 0$, $z_{N-1}\to 1$ and $z_{N}\to \infty$, 
\begin{figure}[h]
      \includegraphics[scale=0.4,angle=0]{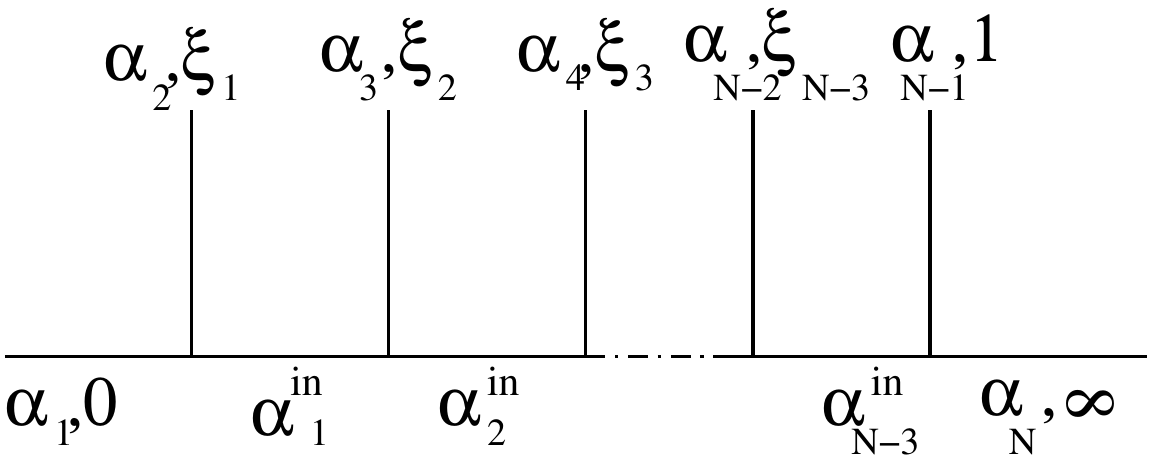}
    \caption{Diagram representing the conformal block (\ref{gencb})}  
   \label{cblockgen}  
   \end{figure}

The  general  conformal block
\begin{equation}
\mathcal{G}(\xi_1,..,\xi_{N-3})\hat{=}\langle e^{\alpha_1 \phi(0)} \prod_{i=2}^{N-2}e^{\alpha_i\phi(\xi_{i-1})}e^{\alpha_{N-1} \phi(1)}e^{\alpha_N \phi(\infty)}\rangle\label{gencb}
\end{equation}
   depends  on $N-3$ variables and it is fully characterized by the $N$ charges $\alpha_i$ of the "external" operators together with  the $N-3$ charges $\alpha_{i}^{in}$ of the internal operators, i.e. the ones appearing  in the fusion channels\cite{diFrancesco}  Consider the conformal block $\mathcal{G}(\xi_1,..,\xi_{N-3})$ represented by the diagram in Fig.({\ref{cblockgen}}) and define $\delta_1=\Delta_{\alpha_{1}^{in}}-\Delta_{\alpha_1}-\Delta_{\alpha_2}$, $\delta_i=\Delta_{\alpha_{i}^{in}}-\Delta_{\alpha^{in}_{i-1}}-\Delta_{\alpha_{i+1}}$ for $i=2,..,N-4$ and $\delta_{N-3}=\Delta_{\alpha_{N}}-\Delta_{\alpha_{N-1}}-\Delta_{\alpha_{N-3}^{in}}$. Using  the following mapping:
\begin{eqnarray}
q_i&=&\xi_{i}/\xi_{i+1} \quad q_{N-3}=\xi_{N-3}\quad  i=1,..,N-4\nonumber \\
\epsilon_1&=&b\quad \epsilon_2=1/b\quad a_i = \alpha^{in}_i - Q/2 (i=1,..,N-3)  \nonumber \\
\mu_1&=&\alpha_2 + \alpha_1-Q/2;\;\; \mu_2 = \alpha_2 - \alpha_1 + Q/2\nonumber \\
\mu_3&=&\alpha_{N-1} + \alpha_{N}-Q/2;\;\; \mu_4 = \alpha_{N-1} - \alpha_{N}+Q/2\nonumber \\
m_i&=& \alpha_{i+2} ( i=0,..,N-1),
\label{mapping_agt}
\end{eqnarray}
the AGT relation states that:
\begin{equation}
\frac{\mathcal{G}(\xi_1,..,\xi_{N-3})}{\prod_{i=1}^{N-3} q_i^{\delta_i}}=Z^{\epsilon_1,\epsilon_2}_{inst}(\vec{{\bf a}},{\bf m},{\bf \mu};{\bf q})Z_{U(1)}^{\epsilon_1,\epsilon_2}({\bf m},{\bf q}),
\label{agt_rel}
\end{equation}
with
\begin{equation}
Z_{U(1)}^{\epsilon_1,\epsilon_2}({\bf m},{\bf q})=\prod_{C_{ij}} (1-q_i q_{i+1}..q_{j})^{2m_i(Q-m_{j+1})}.
\label{zu1}
\end{equation}
In the above formula, the  $C_{ij}$ are all the possible segments connecting consecutive nodes in the diagram (\ref{cblockgen}). The $i$-th node, $i=1,..N-2$,  is associated at the charge $\alpha_{i+1}$ and in (\ref{zu1}) the $m_i=\alpha_{i+2}$ and $m_j=\alpha_{j+2}$ are the charges associated to the $i$-th and the $j(>i)$-th node connected by $C_{ij}$.

{\it AGT relation and Moore-Read wavefunctions} 

\begin{figure}[h]
      \includegraphics[scale=0.4,angle=0]{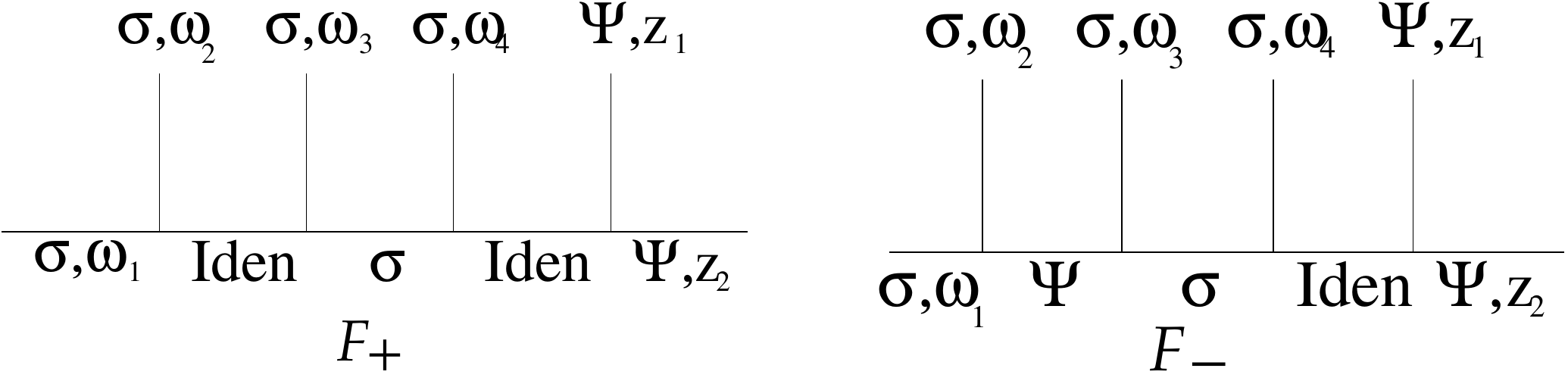}
    \caption{Conformal blocks corresponding to the functions computed in \cite{Nayak_Wilczek}}  
   \label{cblockes}  
   \end{figure}

We now show that the AGT relation can be applied to  the conformal blocks of
the Moore-Read wavefunctions.  
In this respect, an important role is played by the so called  degenerate
fields 
$V_{nm}$ of the Liouville theory  which are defined by:
\begin{equation}
V_{nm} \hat{=}e^{\alpha_{nm} \phi}\quad \alpha_{nm}=\frac{1-n}{2} b +\frac{1-m}{2b} 
\label{degf}
\end{equation} 
where $n$ and $m$ are two integers. 
A degenerate field is a primary of the Virasoro
algebra\cite{diFrancesco},
which presents in the corresponding Verma module a null
vector at level $n m$. 
The degenerate fields form a subset of fields which closes under the operator
algebra. 
For certain values of the parameter $b$
there is within 
the set of degenerate fields a {\it finite} subset which closes under operator
algebra.
 This is what happen for $b=2 i/\sqrt{3}$ (or equivalently $b=-i\sqrt{3}/2$),
 corresponding to $c=1/2$: 
the operator $V_{21}$ and  $V_{12}$, with dimension $\Delta_{21}=1/2$ and $\Delta_{12}=1/16$,  form together with the Identity  the Ising CFT:
\begin{equation}
b=2 i/\sqrt{3}:\quad \Psi \equiv  V_{21} \quad \sigma \equiv  V_{12} . 
\label{assoc}
\end{equation}
 The analytical properties of the Nekrasov partition function with respect to
the charges $\alpha_i$ and $\alpha_{i}^{in}$ (Fig.(\ref{cblockgen})) 
entering in the corresponding conformal
block  have been discussed in
\cite{pogho,Fateev_Litvinov}, showing in particular
the appearance of poles when the
charges $\alpha^{in}_i$ correspond to the  degenerate fields, see (\ref{degf}). 
From this 
we can argue that the AGT relation (\ref{mapping_agt})-(\ref{agt_rel}) is still valid for computing  the conformal block (\ref{qh_cb}) where all operators are degenerate fields.   Indeed, as it has been discussed in \cite{Zamo_recu1} for the Liouville conformal block, the residues at the poles corresponding to degenerate internal operators should vanish  when the external operators are degenerate fields too. This is the reason  why we expect the AGT relation to be valid  also for correlation functions of rational CFTs, and then for quantum Hall wavefunctions.
 To be more concrete, we consider as an illustrative  and non trivial example a six-point conformal block (\ref{qh_cb}) with $n=4$ and $N=2$. These are the 
simplest wavefunctions exhibiting non-Abelian statistics.  There are two possible conformal  blocks $\mathcal{F}_{0,1/2}(w_1,w_2,w_3,w_{4},z_1,z_2)$, represented by the diagram  in Fig(\ref{cblockes}). The explicit expression of these functions  can be found in Eq.(7.16) of \cite{Nayak_Wilczek} and we do not report it here. 
Using a conformal map, we  can send the points $w_1\to 0$, $z_1\to 1$ and $z_2 \to \infty$, which means, in the sphere geometry, that we put one quasi-hole at the South Pole and one particle at the North Pole. We thus study the function
\begin{equation}
\mathcal{F}_{0,1/2}(w_2,w_3,w_{4})\hat{=}\lim_{R\to \infty}R\; \mathcal{F}_{0,1/2}(0,w_2,w_3,w_{4},1,R),
\end{equation}
and we derive its expansions for small $q_i$. This expansion, at the second order, has the following form:
\begin{eqnarray}
\frac{\mathcal{F}_{0}(\overbrace{q_1q_2q_3}^{=w_2},\overbrace{q_2 q_3}^{=w_3},\overbrace{q_3}^{=w_4})}{q_1^{-1/8}q_2^{-1/8}q_3^{-1/4}}&\sim&1+\frac{q_2}{8} +\frac{2q_1^2+9q_2^2+16 q_3^2}{128}+..\nonumber \\ \label{idesp} \\
\frac{\mathcal{F}_{\frac{1}{2}}(q_1q_2q_3,q_2 q_3,q_3)}{q_1^{-3/8}q_2^{1/8}q_3^{1/4}}&\sim&1+\frac{2q_1-3q_2}{8} + \nonumber \\&& \hspace{-2cm} + \frac{18 q_1^2-15 q_2^2+16 q_3^2+20 q_1 q_2}{128} +..\label{pesp}
\end{eqnarray}

In order to check (\ref{agt_rel}), we consider general $b$ and we study the Liouville conformal block (\ref{gencb}) with $N=6$ and:
\begin{equation}
\alpha_i=-\frac{1}{2 b} \,(i=1,..,4)\quad \alpha_5=\alpha_6=-\frac{b}{2}.
\label{extc}
\end{equation}
We are thus considering the Liouville correlator $\langle V_{12}V_{12}V_{12}V_{12}V_{21}V_{21}\rangle$. In this case, there are two possible conformal blocks corresponding to the fusion channels $V_{12} \times V_{12}\to \mbox{Id}$ and $V_{12}\times V_{12}\to V_{13}$. The internal charges appearing in   Fig(\ref{cblockgen}) are then set to
\begin{eqnarray}
\mbox{Id channel:}&&\alpha_{1}^{in}=0\,;\alpha_{2}^{in}=-\frac{1}{2 b}\,;\alpha_{3}^{in}=0\label{idchan}\\
\mbox{$V_{13}$ channel:}&&\alpha_{1}^{in}=-\frac{1}{b}\,;\alpha_{2}^{in}=-\frac{1}{2 b}\,;\alpha_{3}^{in}=0.
\label{pchan}\end{eqnarray}
 For $b \to 2 i\sqrt{3}$ the above  two channels correspond respectively to the conformal blocks $\mathcal{F}_{0}$ and $\mathcal{F}_{1/2}$. Indeed, for $b=2 i\sqrt{3}$, i.e. for $c=1/2$, one can show that the operator $V_{13}$ can be identified with $\Psi$\cite{diFrancesco}.  
 
An alternative route to the computation of the expansions (\ref{idesp}-\ref{pesp})
in terms of the instanton partition function (\ref{zion}) is provided by the dictionary (\ref{mapping_agt}) which determine the set of parameters 
$\vec{{\bf a}},{\bf m},{\bf \mu}$ as a function of $\alpha_i$ and $\alpha_{i}^{in}$. The explicit formulae to compute (\ref{zion}) can be found in \cite{BFMT,FMP} and are resumed in \cite{AGT}. Here we show for instance the case corresponding to   (\ref{idchan}):
\begin{eqnarray}
Z^{b,1/b}_{inst}&\sim& 1 + \frac{3}{2 b^2} q_1 + \left(\frac{3}{2} + \frac{1}{b^2}\right) q_3+\nonumber \\
&&\hspace{-2cm}+\frac{9 + 33 b^2 + 36 b^4 + 12 b^6}{8 b^4 + 12 b^6}q_1^2+\left(2 + \frac{1}{2 b^4} + \frac{2}{b^2}\right)q_3^2+\nonumber \\
&&\hspace{-2.2cm}+\left(1+\frac{3}{2 b^2}\right)q_1q_2 +\left(\frac{3}{2}+\frac{1}{b^2}\right)q_2q_3+\frac{6 + 13 b^2 + 6 b^4}{4 b^4}q_1q_3+..\nonumber \\
\end{eqnarray}
 Taking into account the function (\ref{zu1}), that in the case under consideration takes the form:
 \begin{eqnarray}
 Z^{b,1/b}_{U(1)}(q_1,q_2,q_3)&=&\left[(1-q_1)(1-q_2)(1-q_1q_2)\right]^{1+\frac{3}{2 b^2}}\nonumber \\&&\left[(1-q_3)(1-q_2q_3)(1-q_1q_2q_3)\right]^{\frac{3}{2} +\frac{1}{b^2}} \nonumber
 \end{eqnarray}
   we find, by setting  $b\to 2i/\sqrt{3}$,   the expansion  
(\ref{idesp}). The same can be shown also for the expansion (\ref{pesp}). The expansions (\ref{idesp}-\ref{pesp}) can then be directly related to the one given in 
(\ref{zion}).

In this Letter we have shown a neat connection between the theory of the Moore-Read non-Abelian quantum Hall states and four dimensional superconformal $SU(2)$ quiver gauge theories. 
We provided a mathematical interpretation linking the Moore-Read wavefunctions 
to generating functionals of ${\mathbb Z}_n$-equivariant Donaldson polynomials. Let us notice that our analysis
unveil some special properties of the submanifold (\ref{degf}) of the equivariant parameter space. Indeed, it is well known that correlation functions of $V_{nm}$ degenerate fields satisfy order $nm$ partial differential equations: it would be interesting to investigate the interpretation of this
result in Donaldson theory, trying to make a connection with the results of \cite{braverman}.
   
Our result opens a bridge between different theories that can provide new insights in their comprehension; nonetheless, the origin
of this relation is still mostly unclear.
Some arguments for the AGT correspondence have been proposed in the context of topological strings and M-theory \cite{DV,BT,NW}. 
One line of arguments that could be relevant for our problem is that  
the four dimensional gauge theories in question can be realized as superstring compactifications in presence of D-branes.
In particular \cite{DV} one can consider B model topological branes in the local geometry of a resolved $A_1$ singularity fibered over a Riemann surface
$\Sigma$ \footnote{Although in this letter we focused just on the sphere, our arguments
apply to any Riemann surface.} .  
The dynamics of such branes is governed by a holomorphic Chern-Simons theory, which upon reduction
to the blown-up ${\mathbb P}^1$ provides a matrix model with a $\beta$-ensemble measure induced by the non-triviality of the fibration. 
The collective field describing the large $N$ limit of this matrix model is precisely a Liouville field 
on the double covering of $\Sigma$
\cite{russi,DV}! It is thus tempting to suppose that this matrix model captures 
at least some aspects of the microscopic description of the non-Abelian Fractional Quantum Hall systems
\footnote{Different earlier proposals of matrix quantum mechanical models inspired by D0-branes appeared in L.~Susskind,
  arXiv:hep-th/0101029, for a review see A.~Cappelli and I.~D.~Rodriguez,
  J.\ Phys.\ A  {\bf 42} (2009) 304006.}.
 


There are several interesting aspects that are worth to be investigated further.
 For instance, it would be interesting to study the consequences of  the vanishing of the Berry connection, proven for Moore-Read states\cite{Read_holonomy},  in terms of $SU(2)$ quiver gauge theories. As a final remark, we mention that AGT relation (\ref{agt_rel}) can be extended to a relation  between the $SU(k)$ quiver gauge theories and $WA_{k-1}$ Toda theories \cite{Toda}. The results presented in this Letter  are expected to  be generalized  to  the states based to the affine $WA_{k-1}$ Toda theories\cite{Jack}. These states include as a special case the Moore-Read states and their  most direct generalization, the so-called $Z_k$ Read-Rezayi states \cite{ReadRezayi}.  The ground state wavefunctions of these states can be written as particular Jack polynomials \cite{Jack}. Via the AGT relation, these Jack  polynomials are related to Nekrasov partition functions. It is compelling to gain further insights on the origin of all  these relations. 

{\it Acknowledgements}: The authors thanks  G.~Bonelli, U.~Bruzzo, and R.~Poghossian for very helpful discussions. 
They also acknowledge  conversations with A.~Cappelli and  B.~Estienne.

\end{document}